\documentclass{emulateapj}
\lefthead{FUENTES ET AL.} \righthead{HYPERVELOCITY STAR'S VARIABILITY}
\def\kms{{~\rm km\,s^{-1}}}
\def\eq#1{equation (\ref{#1})}

\def\kpc{\rm kpc}

\def\days{{~\rm days}}
\def\msun{M_\odot}

\def\lsun{L_\odot}
\def\teff{T_{\rm eff}}

\begin{document}

\title{The Hypervelocity Star SDSS J090745.0+024507 is a Short-Period
Variable\altaffilmark{1}}

\author{Cesar I.\,Fuentes\altaffilmark{2},
K.\,Z.\,Stanek\altaffilmark{2,3},
B.\,Scott\,Gaudi\altaffilmark{2},
Brian\,A.\,McLeod\altaffilmark{2},
Slavko\,Bogdanov\altaffilmark{2},
Joel\,D.\,Hartman\altaffilmark{2},
Ryan\, C.\,Hickox\altaffilmark{2},
Matthew\,J.\,Holman\altaffilmark{2}}

\altaffiltext{1}{Based on data from the MMTO 6.5m telescope
and the FLWO 1.2m telescope}

\altaffiltext{2}{Harvard-Smithsonian Center for Astrophysics, 60
Garden Street, Cambridge, MA 02138;  cfuentes@cfa.harvard.edu}

\altaffiltext{3}{Department of Astronomy, The Ohio
State University, Columbus, OH 43210}

\begin{abstract}

We present high-precision photometry of the hypervelocity star SDSS
J090745.0+024507 (HVS), which has a Galactic rest-frame radial
velocity of $v=709\kms$, and so has likely been ejected from the
supermassive black hole in the Galactic center.  Our data were
obtained on two nights using the MMT 6.5m telescope, and is
supplemented by lower precision photometry obtained on four nights
using the FLWO 1.2m telescope.  The high-precision photometry
indicates that the HVS is a short-period, low-amplitude variable, with
period $P=0.2-2\days$ and amplitude $A = 2-10\%$.  Together with the
known effective temperature of $\teff\simeq 10,500~{\rm K}$ (spectral
type B9), this variability implies that the HVS is a member of the
class of slowly pulsating B-type main sequence stars, thus resolving
the previously-reported two-fold degeneracy in the luminosity and
distance of the star.  The HVS has a heliocentric distance of
$71~\kpc$, and an age of $\la 0.35~{\rm Gyr}$.  The time of ejection
from the center of the Galaxy is $\le 100$ Myr, and thus the existence
of the OS constitutes observational evidence of a population of young
stars in the proximity of the central supermassive black hole $\sim
0.1$ Gyr ago.  It is possible that the HVS was a member of a binary
that was tidally disrupted by the central black hole; we discuss
constraints on the properties of the companion's orbit.
\end{abstract}

\section{Introduction}\label{sec:intro}

In their survey of blue horizontal branch (BHB) stars, \citet{brown05}
discovered that the Galactic halo star SDSS J090745.0+024507
(hereafter HVS) has a radial velocity of $853 \pm 12 \kms$.  This
corresponds to a Galactic rest-frame radial velocity of $\sim 700
\kms$, well above the local escape speed from the Galaxy.  As reviewed
by \citet{brown05}, the only plausible mechanism for achieving this
extreme velocity is ejection from the vicinity of the central
supermassive black hole (CBH), as predicted by \citet{hills88}, and
studied in detail by \citet{yu03}.  The detection of the HVS and other
hypervelocity stars is important because it allows one to probe the
population of stars near the CBH in the Milky Way's recent past.  In
addition, it may be possible to use these stars to constrain the
various scenarios for the origin of the young stars detected near the
Galactic center \citep{genzel03,ghez05,gould03,hansen03}.  Finally,
precise proper motion measurements of hypervelocity stars ejected from
the Galactic center can be used to probe the shape of the Galactic
halo \citep{gnedin05}.

The intrinsic luminosity and distance to the HVS suffers from a
two-fold degeneracy which hampers the interpretation of its origin.
This degeneracy arises from the coincidence that the main sequence and
horizontal branch overlap at the measured effective temperature
$\teff=10,500~{\rm K}$ and surface gravity of the HVS.  Thus the HVS
could be either a BHB giant or a B9 main-sequence star.  The intrinsic
luminosities of these two types of stars differ by a factor of $\sim
4$, and thus the inferred distances to the HVS differ by a factor of
$\sim 2$.  The distance is $39~\kpc$ or $71~\kpc$ if the HVS is a BHB
star or main-sequence star, respectively.

In order to search for photometric variability and so pin down its
properties, we performed precise photometry of the HVS using the MMT
6.5m telescope.  We describe our observations and data reduction in
\S\ref{sec:data}.  As we discuss in \S\ref{sec:ana}, we find that the
star is indeed variable, and we constrain the variability to be
short-period ($0.2-2\days$) and low-amplitude ($2-10\%$).  Together
with the measured effective temperature, we argue in \S\ref{sec:spb}
that this puts the HVS in the class of B9 main-sequence stars that
pulsate with periods of the order of one day, so-called slowly
pulsating B (SPB) stars \citep{waelkens91}.  Therefore, our
observations resolve the two-fold ambiguity and indicate that the HVS
is a main-sequence star with heliocentric distance of $71~\kpc$.  We
discuss the implications of this result in \S\ref{sec:disc}.

\section{Observations and Data Reduction}\label{sec:data}

We obtained photometric data on the HVS during two nights (UT 2005
January 15, April 13) with the Megacam CCD camera \citep{mcleod00} on
the MMT 6.5m telescope, and during four nights (UT 2005 January
13,15-17) with the 1.2-meter telescope at FLWO.  Using MegaCam we
obtained 9 $g$-band images on the first night, and 16 $g$-band and 9
$r$-band images on the second night.  These high signal-to-noise ratio
data were used to detect the variability of the star. The FLWO data
were obtained in the $V$-band over 7 epochs, in order to constrain the
amplitude of the variability.

The raw images were reduced in the usual manner. Photometry was
carried out using PSF-fitting photometry with the package DAOPHOT II
\citep{stetson87,stetson92}. We used 15-30 reference stars to obtain
the relative photometry between different epochs.  We roughly
calibrated our relative photometry using Sloan Digital Sky Survey
(SDSS) absolute photometry \citep{abazajian05} of the nearby
comparison star SDSS J090751.07+024534.9.

\begin{figure}
\epsscale{1.0}
\plotone{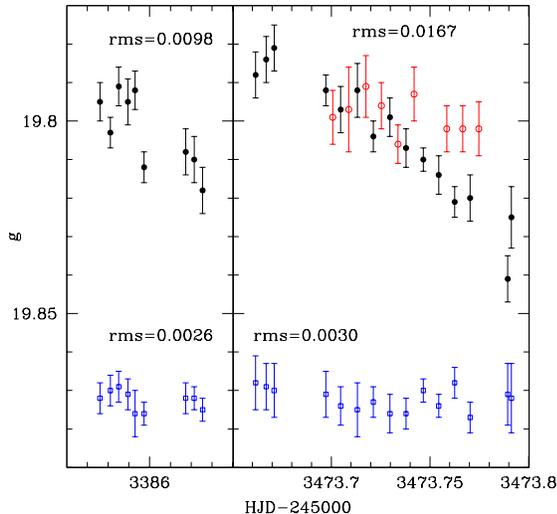}
\caption{\label{fig:curve}
Relative photometry of the hypervelocity star SDSS
J090745.0+024507 (HVS). The black solid circles show the $g$-band
magnitude of the HVS versus HJD-2450000. In the second panel the red open
circles show the $r$-band data.  The blue open squares show
the photometry for the slightly fainter and nearby star
SDSS J090751.07+024534.9.  Note that the scales in both panels are the same.
}
\end{figure}

Figure \ref{fig:curve} shows the $g$-band and $r$-band light curves
for the HVS constructed from the MegaCam data. The $g$-band data from
the two nights are shown in separate panels.  The HVS is variable in
the $g$-band, but shows no evidence for variability within the errors
in the $r$-band.  The $g$-band variability is certainly real, as
evidenced by the photometry of the slightly fainter and nearby
comparison star SDSS J090751.07+024534.9, which is constant to within
the errors (see Fig.\ \ref{fig:curve}).

The FLWO data are of considerably lower precision than the MMT data,
with photometric errors of $\sim 5-10\%$, as compared to $\sim
0.5-1\%$ for the MMT data.  The FLWO data therefore do not provide
strong constraints on the periodicity of the HVS's variability.  As we
explain in \S\ref{sec:ana}, these data are nevertheless very useful
because they constrain the amplitude of the variability to be $\la
10\%$.

\section{Analysis and Results}\label{sec:ana}

Figure \ref{fig:curve} shows that the HVS star is variable in the
$g$-band. The RMS deviation over the $\sim$ 3.5 hours of data on the
second night is nearly 6 times larger than that of the comparison
star. A constant flux fit to the $g$-band data has $\chi^2=155.9$ for
$24$ degrees-of-freedom (dof), and so is a poor representation of the
data. On the other hand, the $r$-band data are consistent with a
constant flux, with $\chi^2=4.5$ for $8$ dof.

In order to constrain the period and amplitude of the variability, we
fit the data to the model:
\begin{equation}
m(t_i)=A\sin\left[\frac{2\pi}{P}(t_i-t_0)+\phi_0\right]+m_{0,j},
\label{eqn:model}
\end{equation}
where $m(t_i)$ is the magnitude of observation $i$ 
taken at time $t_i$,  $P$ is the period, $A$ is the amplitude, 
and $m_{0,j}$ and $\phi_0$ correspond to the magnitude zero point 
and phase at the error-weighted mean
observation time, $t_0-2450000.=3441.95357$.  We assume separate
magnitude zero points $m_{0,j}$ for each data set $j$ (MMT or FLWO).  
Fitting \eq{eqn:model} is equivalent to
a Lomb-Scargle periodogram with a floating mean
\citep{lomb76,scargle82,cumming04}.   We  search for fits at $10^6$
equally-spaced steps in $\log P$ in the range 
$-2\le \log(P/{\rm days})\le 2$.

If we fit only the MMT $g$-band data, we can only place a lower limit
on the period and amplitude of the variability, $A \ge 2\%$ and $P \ge
0.2\days$.  This fact can be understood from inspection of the data in
Figure \ref{fig:curve}.  The $g$-band data for both nights are nearly
consistent with a simple linear decline with slope $m \sim 0.4~{\rm
mag/day}$.  Thus the data can be approximately described by any model
satisfying $A \simeq (2\pi)^{-1} mP$, for $P$ much larger than the
duration of the observations on any given night, $\sim 3.5~{\rm hr}$.

\begin{figure}
\epsscale{1.0} 
\plotone{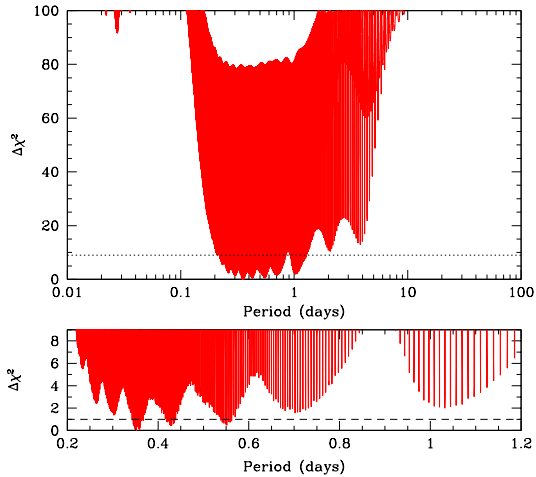}
\caption{\label{fig:two} The difference in $\chi^2$ of a sinusoidal
model fit to the HVS light curve from the minimum $\chi^2$ of the
best-fit model with $P=0.355188\days$, as a function of the period of
the model.  The upper panel shows the full range of periods searched
and the lower is a zoom in the regime of most interest.  }
\end{figure}

\begin{figure}
\epsscale{1.0} 
\plotone{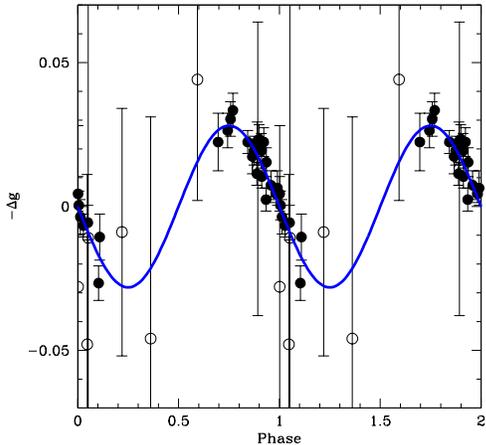}
\caption{\label{fig:three} The photometry of the HVS is displayed as a
function of phase angle, folded according to the best fit period
$P=0.355188\days$ and with the zero point subtracted. The black points
are the MMT $g$-band photometry for the HVS. The open circles are the
1.2-meter $V$-band photometry.  Points are plotted twice for clarity.
}
\end{figure}

Fitting both the $g$-band MMT and $V$-band FLWO datasets
simultaneously constrains the period and amplitude of the variability.
We assume that the amplitude, period, and phase of the variability is
the same in the $g$ and $V$ bands.  Figure \ref{fig:two} shows the
resulting periodogram, plotted as $\Delta\chi^2\equiv
\chi^2-\chi^2_{\rm min}$ versus $P$.  We find a best-fit for $P=
0.355188 \pm 0.000021 \days$, $A= 0.0280\pm 0.0033 ~{\rm mag}$, and
$\phi_0=2.98 \pm 0.17$, with $\chi^2_{\rm min}=24.5$ for 32-5=27 dof.
The best-fit model is shown in Figure \ref{fig:three}, together with
both the MMT and FLWO data folded about the best-fit period.  This fit
is not unique.  There are flanking aliases with periods separated by
$2.06~{\rm min}$, corresponding to an integer number of additional
cycles between the $\sim 88\days$ separating the two nights of the
$g$-band data.  There are also fits at periods of $\sim 0.43\days$ and
$\sim 0.55\days$ that are equally good ($\Delta\chi^2\le 1$).
Finally, essentially all periods with $P=0.2-1.5\days$ are allowed at
the $3\sigma$ level.

\begin{figure}
\epsscale{1.0} 
\plotone{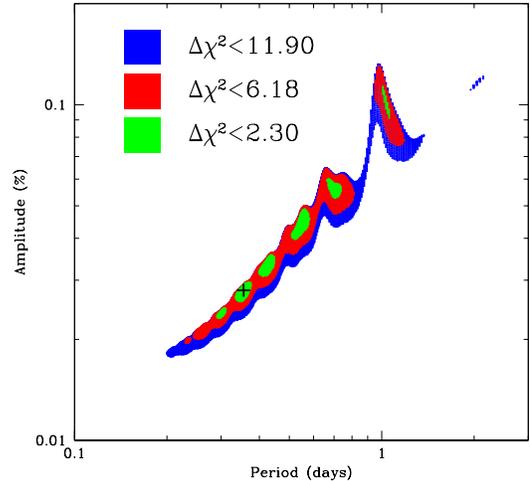}
\caption{\label{fig:zero} The 1, 2, and $3\sigma$ allowed regions
($\Delta \chi^2 \le 2.30, 6.18, 11.90$) in the amplitude-period
$(A,P)$ plane for a sinusoidal fit to the HVS light curve.  The best
fit is attained with $P=0.355188 \pm 0.000021\days$ and $A=0.0280\pm
0.0033$~mag, and is indicated by the cross.  }
\end{figure}

Because the $g$-band data on any individual night show little
curvature, only the slope is well constrained, and so the amplitude of
the fit is correlated with the period.  This is shown in Figure
\ref{fig:zero}, where we show the 1, 2, and 3$\sigma$ allowed regions
in the $A-P$ plane.  We see that longer periods require larger
amplitudes.  However, regardless of the best fit, we can rule out
periods $P\la 0.2\days$ and $\ga 1.5\days$ and amplitudes $A\la 2\%$
and $\ga 10\%$.

The $r$-band data show no evidence for variability.  Adopting the
best-fit period and phase from the $g$ and $V$-band data, the
amplitude of the $r$-band variability is $A_r \le 0.016~{\rm mag}$ at
the $95\%$ confidence level.

\section{The Hypervelocity Star is a Slowly Pulsating B-type Main-Sequence Star}\label{sec:spb}

The amplitude and period of the HVS's variability, as well as its
effective temperature, are all consistent with the class of slowly
pulsating B-type main-sequence stars first identified by
\citet{waelkens91}.  These stars are multi-periodic, non-radial
pulsators with periods of $P=0.4-4$ days, amplitudes of a few
millimagnitudes to a few percent, and effective temperatures of
$T_{\rm eff}=10,000-20,000~{\rm K}$.  According to \cite{waelkens98}
the HVS would fall in the low temperature boundary of the instability
strip calculated by \cite{pamyatnykh99} for SPBs.  In fact, there is a
striking similarity between the HVS and the known SPB star HD45953,
reported by \cite{waelkens98}, which is also at the low temperature
edge of the class with $T_{\rm eff}=11,500$ K, $P=0.43$ days, and
$M=2.89\pm0.07 \msun$ \citep{decat02}.  SPB stars are also known to
have lower amplitudes in redder bandpasses, with relative amplitudes
that are consistent with the upper limit on the HVS's $r$-band
variability of $0.016~{\rm mag}$.

On the other hand, there is evidence that BHB stars do not vary at
this level. Studying a predominantly BHB cluster with high precision
photometry, \citet{contreras05} found more than 200 RR Lyrae variables
(with similar periods to the HVS), but no BHB stars that showed
significant variation.  See \citet{catelan05} for more discussion on
the variability of HB stars.

The fact that the HVS is variable with $A=2-10\%$, and that BHB stars
do not vary at this level, implies that it is a main sequence
star. Hence its temperature $T_{\rm eff} \sim 10,500 K$ indicates a
luminosity of $L\sim 160\lsun$, giving a distance of 71 kpc
\citep{brown05}. Using the reported velocity for the HVS ($709\kms$)
and assuming its movement is only radial we obtain a travel time from
the center of the galaxy of $\le 0.1$ Gyr. For a $3\msun$ B9 star, the
evolutionary tracks of \cite{schaller92} give a main-sequence lifetime
of $0.35$ Gyr.

\section{Summary and Discussion}\label{sec:disc}

We have performed high-precision, time-series photometry of the
hypervelocity star SDSS J090745.0+024507, which shows that it is a
low-amplitude, short-period variable. A sinusoidal fit to the light
curve yields a best-fit period of $P \simeq 0.36\days$ and amplitude
of $A \simeq 2.5\%$, however this fit is not unique and the exact
values of the amplitude and period of the variation are poorly
constrained.  Nevertheless, the period and amplitude are constrained
to be in the range $A=2-10\%$ and $P=0.2-1.5\days$.

Together with the known effective temperature of $\teff\simeq
10,500~{\rm K}$ (spectral type B9), this variability implies that the
HVS is a member of the class of slowly pulsating B-type main sequence
stars identified by \cite{waelkens91}.  This identification resolves
the previously-reported two-fold degeneracy in the luminosity and
hence distance to the HVS.  The HVS has a mass of $\sim 3M_\odot$, an
age of $\la 0.35~{\rm Gyr}$, a heliocentric distance of $\sim 71$ kpc
and a travel time from the Galactic center $\sim 0.1$ Gyr.

The HVS can be used to probe the population of stars near the CBH in
the Milky Way's recent past.  The most plausible mechanism for
creating hyper-velocity stars such as the HVS is a strong
gravitational interaction with the Milky Way's CBH, perhaps as a
member of a short-period binary that was disrupted by the tidal field
of the CBH \citep{hills88,yu03} As such, the existence of the HVS
implies that young stars must have been present within $\sim 0.1~{\rm
pc}$ of the CBH $\sim 0.1$ Gyr ago.  There is observational evidence
for even younger stars with ages of $\sim 1-10~{\rm Myr}$ currently
orbiting the CBH \citep{ghez03,genzel03} at distances of $\la 0.1~{\rm
pc}$.  The existence of such stars is puzzling, as strong tidal
interaction with the black hole should prevent local star formation
(but see \citealt{levin03} and \citealt{milos04}), and yet these stars
are unlikely to live long enough to be scattered into such close
orbits.  Thus the existence of young stars near the CBH today
\citep{ghez03,genzel03}, along with the inference from the HVS that
they were also present $\sim 0.1$ Gyr ago, may suggest that the
mechanism for delivering young stars to the CBH must be efficient and
continuous, which may in turn constrain the various models for such
delivery \citep{genzel03,ghez05,gould03,hansen03}.  If the HVS
migrated to near the CBH from outside, its expected MS lifetime
($\sim0.35$ Gyr), implies the migration time must be $\la 0.2$ Gyr.

Assuming that the HVS is the ejected component of a binary that was
tidally disrupted by the CBH, we can use its known mass and ejection
velocity to place constraints on the properties of original binary,
and on companion's current orbit about the CBH (see also
\citealt{gualandris05}). Assuming a mass for the companion $m_1$, we
can determine that initial separation of the binary $a_{bin}$, and the
peribothron $q$ and eccentricity $e$ of the companion's orbit around
the CBH \citep{gould03,yu03}.  We find that the eccentricity of the
bound star's orbit is always high, ranging from $e=0.97$ to $e\sim 1$
as an increasing function of $m_1$. The binary separation $a_{bin}$
and $q$ increase with $m_1$. For a companion with $m_1=3~M_{\odot}$
(i.e.\ an equal-mass binary), we find $a_{bin}=0.69$ AU, $q$=72.8 AU
and $e$=0.98.  The semimajor axis of the bound star's orbit about the
CBH is $\sim 3830$ AU.

The period of the companion of the HVS about the CBH will be of the
order of $\sim 100$ yr.  Furthermore, a $\sim 3~M_\odot$ main-sequence
star at the Galactic center would have $K \simeq 18.5$, over a
magnitude fainter than the faintest stars for which accurate orbits
are being measured in near-IR imaging studies of the Galactic center
using $10{\rm m}$-class telescopes \citep{schodel03,ghez05}.  As a
result, unless it happens to be near peribothron and massive, the
companion to the HVS will probably be difficult to find.  Assuming
tidal disruption of binaries is the mechanism by which most HVS are
formed, it may be difficult to link an observed HVS with its orbiting
stellar companion.

The fact that the HVS is an SPB star implies that it should be
possible to study its properties in more detail with follow-up
observations using asteroseismology.  SPB stars show multi-periodic
variability, with secondary period amplitudes that are of the same
order as that of the primary period.  This variability is thought to
be due to non-radial g-mode pulsations \citep{waelkens91}.  Hence, the
HVS should be targeted for more photometric data in order to unravel
the possible different pulsation modes. Matching these g-modes would
give additional and precise information about the properties of the
star. We are specially interested in HVS's age, since an independent
measurement from asteroseismology would provide further evidence of
its origin and history across the Galaxy.

\acknowledgments We would like to thank Andy Gould for helpful
discussions, Warren Brown, Margaret Geller, and Scott Kenyon
for useful comments on the manuscript, and Barbara Mochejska for
help with the SM plotting package. BSG was supported by a Menzel
Fellowship from the Harvard College Observatory.

\end{document}